\documentclass{emulateapj}

\newcommand{\micr}{$\,\mu{\rm m}$}
\newcommand{\ptune}{$P_{\rm tune}$}
\newcommand{\lmm}{lines ${\rm mm}^{-1}$}
\newcommand{\s}{{\it s}}
\newcommand{\p}{{\it p}}
\newcommand{\spphased}{{\it s}-{\it p}-phased}

\begin{document}

\title{Volume Phase Holographic Gratings: \\
  Polarization Properties and Diffraction Efficiency}
\shorttitle{Volume Phase Holographic Gratings}
\author{I.~K.~Baldry$^{1,2}$, J.~Bland-Hawthorn$^{2}$ and 
  J.~G.~Robertson$^{2,3}$}
\affil{$^1$Department of Physics \& Astronomy, Johns Hopkins University, 
  Baltimore, MD~21218, USA (E-mail: baldry@pha.jhu.edu)\\
$^2$Anglo-Australian Observatory, P.O.~Box~296, Epping, NSW~1710, Australia\\
$^3$School of Physics, University of Sydney, NSW~2006, Australia}
\shortauthors{Baldry, Bland-Hawthorn \& Robertson}
\slugcomment{2004 February 16, accepted by Publ.~Astron.~Soc.~Pacific}

\begin{abstract}
  We discuss the polarization properties and first-order diffraction
  efficiencies of volume phase holographic (VPH) transmission gratings, which
  can be exploited to improve the throughput of modern spectrographs.  The
  wavelength of peak efficiency can be tuned by adjustment of the incidence
  angle. We show that the variation of the Kogelnik efficiency versus Bragg
  angle depends only on one parameter, given by $P_{\rm tune} = (\Delta n \,
  d)/(n \, \Lambda)$, where: $\Delta n$ is semi-amplitude of the refractive
  index modulation; $n$ is the average index; $d$ is the thickness of the
  active layer; and $\Lambda$ is the grating period.  The efficiency has a
  well defined dependence on polarization. In particular, it is possible to
  obtain theoretical 100\% diffraction efficiency with one linear polarization
  at any angle or to obtain 100\% efficiency with unpolarized light at
  specific angles.  In the latter case, high efficiency is the result of
  aligning the peaks of the $s$- and $p$-polarization
  efficiency-versus-thickness curves.  The first of these `\spphased\ 
  gratings' for astronomy is in use with the 6dF spectrograph.  Consideration
  of polarization is particularly important for high spectral resolution,
  which requires large incidence angles.  We also discuss the possibility of
  separating polarization states for improved throughput along the entire
  optical train of a spectrograph.
\end{abstract}

\section{Introduction}
\label{sec:intro}

Astronomical spectrographs have undergone a major revolution during the past
few decades \citep{vBBH00,IM00,IM03,LM02,AD03}.  The revolution has
concentrated on the multiplex advantage in order to allow large numbers of
objects or contiguous spatial elements to be observed simultaneously.  This is
possible because large detectors are now available, which can also lead to
wide angle fields and/or wide spectral coverage.

Even though modern spectrographs can achieve up to 40\% throughput
(optics$+$disperser$+$detector), instrumental throughput remains a key issue
for spectrograph design. Moreover, some fraction of the light lost along the
optical train is received as stray light at the detector, and usually provides
a major source of systematic error in the detected signal.

Now that detectors are widely available with 90\% quantum efficiency in the
visible wavelength region, the remaining gains must come from more efficient
designs of the optics and dispersing element or elements, which we discuss.
We concentrate specifically on volume phase holographic (VPH) gratings used in
transmission \citep{Arns95}. However, we note that similar consideration could
apply to a much wider class of dispersing elements, e.g., reflection gratings,
prisms.

What has not been discussed widely is the advantages of the polarization
properties of VPH gratings, in particular, achieving the ideal of 100\%
throughput at any diffraction angle in one linear polarization. In addition, a
theoretical diffraction efficiency of 100\%, in both polarizations, can be
achieved at specific angles with particular instrument configurations. An
instrument that exploits the advantages of VPH gratings can, in principle,
greatly reduce systematic error in the detected signal.

VPH gratings are already in use or are being brought into use in a number of
spectrographs, including: 
LDSS++ and Taurus at the Anglo-Australian Telescope \citep{glazebrook98jan}; 
OSIRIS at the Gran Telescopio Canarias \citep{cepa00};
Goodman spectrograph at the SOAR Telescope \citep*{CES00};
M2HES at the Magellan~II Telescope \citep{bernstein02};
FORS at the Very Large Telescope \citep*{MDR02}; 
FOCAS at the Subaru Telescope \citep{ebizuka03}; 
and LRS at the Hobby-Eberly Telescope \citep{hill03}. 
The potential of VPH gratings for astronomical applications has been
investigated and discussed by Barden and others (\citealt*{BAC98};
\citealt{barden00,barden00proc,barden02}; \citealt*{BCY03};
\citealt{robertson00}; \citealt*{RRD03}; \citealt{tamura04}).  Here, we
elucidate the physics of VPH gratings with emphasis on their polarization and
tuning properties, and we discuss how these properties might be exploited to
improve the performance of spectrographs. In \S~\ref{sec:intro-vph}, we
describe the physics of VPH transmission gratings; in
\S~\ref{sec:example-cases}, we describe some applications taking advantage of
the well-defined polarization properties; in \S~\ref{sec:summary}, we
summarize; and in the Appendix, we give equations for calculating the
resolving powers of transmission gratings immersed between prisms.

\section{VPH Grating Physics}
\label{sec:intro-vph}

In a VPH transmission grating, light is diffracted as it passes through a thin
layer (3--30\micr) of, typically, `di-chromated gelatin' (DCG)
\citep{Shankoff68,Meyerhofer77,Rallison92} in which the refractive index is
modulated approximately sinusoidally.  The modulations are produced by the
interference of two large collimated laser beams, and subsequent processing.
These gratings offer a number of advantages over other gratings, including the
following:
\begin{enumerate} 
\item Diffraction efficiencies can approach 100\% near the design wavelength.
\item The wavelength of peak efficiency can be tuned by adjustment of the
  incidence angle.
\item The line density can be significantly higher (up to 6000 \lmm) than the
  maximum generally available for ruled gratings, which is about 1200 \lmm.
\item Transmission gratings allow shorter pupil relief between the grating and
  both the collimator and camera, which can reduce the required camera
  aperture, increase the field of view and/or improve the point spread
  function (PSF).
\item The grating is sandwiched between glass substrates providing a robust
  device, which can be easily cleaned and have anti-reflection (AR) coatings
  applied.
\item Large grating sizes are feasible.
\end{enumerate}
Further advantages and disadvantages are described by \citet{barden00}.  In
this paper, we also consider the ability to optimize the efficiency for a
particular polarization state.

\subsection{Diffraction by a VPH grating}
\label{sec:diffrac-vph}

Light passing through a VPH grating obeys the usual grating equation,
given by
\begin{equation}
 \frac{m \lambda}{n_i} = \Lambda_g (\sin \alpha_i + \sin \beta_i)
 \label{eqn:grating}
\end{equation}
where: $m$ is an integer (the spectral order); $\lambda$ is the wavelength in
vacuum; $n_i$ is the refractive index of the medium; $\Lambda_g$ is the
grating period, which is the projected separation between the fringes in the
plane of the grating, equivalent to the groove spacing on a ruled grating;
and $\alpha_i$ is the angle of incidence and $\beta_i$ is the angle of
diffraction from the grating normal (the sign convention is such that $\beta_i
= - \alpha_i$ means no diffraction, i.e., zeroth order).  Note that the
grating equation can apply to angles in the DCG layer ($i=2$), in the glass
substrates ($i=1$) or in the air ($i=0$) as long as the air-glass boundaries
are parallel to the DCG layer (see the Appendix for the general case).
Figure~\ref{fig:grating} shows a diagram of a VPH grating with the appropriate
angles and lengths defined.

\begin{figure}
\epsscale{1.15}
\plotone{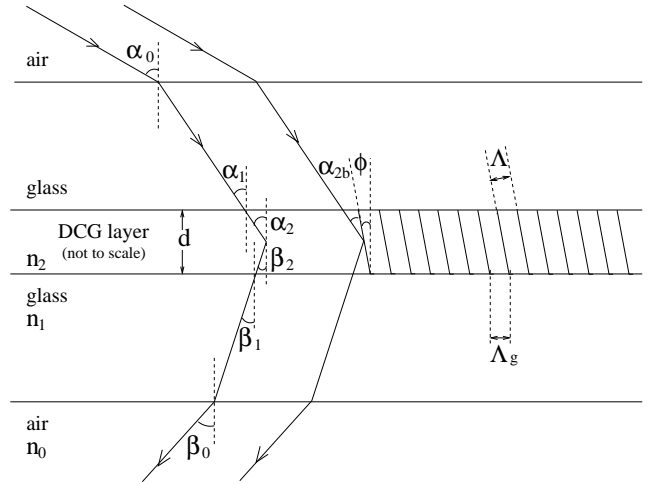}
\caption{Diagram of a VPH grating.{}  
  The equally-spaced lines in the DCG layer represent the peaks of a modulated
  refractive index ($n_2$ is the average value).  Typically, $n_2$ is in the
  range 1.2--1.5 depending on the DCG processing \citep{RS92}, $n_1\simeq1.5$
  and $n_0=1$.  For unslanted fringes, $\phi=0\degr$, $\Lambda=\Lambda_g$ and
  $\alpha_{2b}=\alpha_2$.}
\label{fig:grating}
\end{figure}

For the simplest VPH transmission grating, the plane of the fringes is
perpendicular to the plane of the grating (we use the term `unslanted
fringes').  In this case, $\Lambda_g$ is the same as the separation between
the fringes $\Lambda$.  For the general case,
\begin{equation}
 \Lambda_g = \frac{\Lambda}{\cos \phi}
\end{equation}
where $\phi$ is the `slant' angle between the grating normal and the plane of
the fringes. 

\subsection{The Bragg condition}
\label{sec:bragg-angle}

In a VPH grating, high diffraction efficiency can occur when the light is
effectively `reflected' from the plane of the fringes, i.e.,
\begin{equation}
 \beta_2 + \phi = \alpha_2 - \phi
 \label{eqn:reflec}
\end{equation}
where $\alpha_2$ is the angle of incidence and $\beta_2$ is the angle of
diffraction from the grating normal in the DCG layer.  The phenomenon is
analogous to Bragg `reflection' of X-rays from the atomic layers within a
crystal lattice. In both cases the thickness of the medium being $\gg\lambda$
can result in constructive interference of scattered radiation in that
direction.  The essential role of the non-zero thickness of the DCG layer is
responsible for the term `volume' in VPH gratings.  This `reflection' combined
with the grating equation gives the well known `Bragg condition', which can be
written as
\begin{equation}
 \frac{m \lambda}{n_2} = 2 \Lambda \sin \alpha_{2b}
 \label{eqn:bragg}
\end{equation}
where $n_2$ is the refractive index of the DCG layer and $\alpha_{2b}$ is the
angle of incidence with respect to the plane of the fringes, i.e.,
$\alpha_{2b} = \alpha_2 - \phi$.  Under this condition, $\alpha_{2b}$ is
called the {\em Bragg angle}.  Light nearly obeying this condition is still
diffracted according to the grating equation (Eqn.~\ref{eqn:grating}) but
usually with lower efficiency.  At wavelengths or angles sufficiently outside
the Bragg condition, light passes through the grating without being
diffracted.  The Bragg angle is an important parameter for diffraction by VPH
gratings.  It directly affects efficiency and bandwidth
(\S\S~\ref{sec:eff-1st-ord}--\ref{sec:bandwidth}) and indirectly affects
resolving power (Appendix).

We note that unslanted fringes may be preferred because, with slanted fringes,
the tilt may change or the fringes may curve during DCG processing
\citep{RS92}.  For unslanted fringes ($\phi =0$, $\Lambda=\Lambda_g$, $n_2
\sin \alpha_{2b} = n_i \sin \alpha_i$), the Bragg condition can also be
written as
\begin{equation}
 \frac{m \lambda}{n_i} = 2 \Lambda_g \sin \alpha_i \:.
 \label{eqn:bragg-littrow}
\end{equation}
This defines the Bragg wavelength for a given order of diffraction $m$, and
corresponds to {\em Littrow} diffraction because $\beta_i = \alpha_i$.  

\subsection{First-order diffraction efficiencies}
\label{sec:eff-1st-ord}

\begin{figure}
\epsscale{2.30}
\plottwo{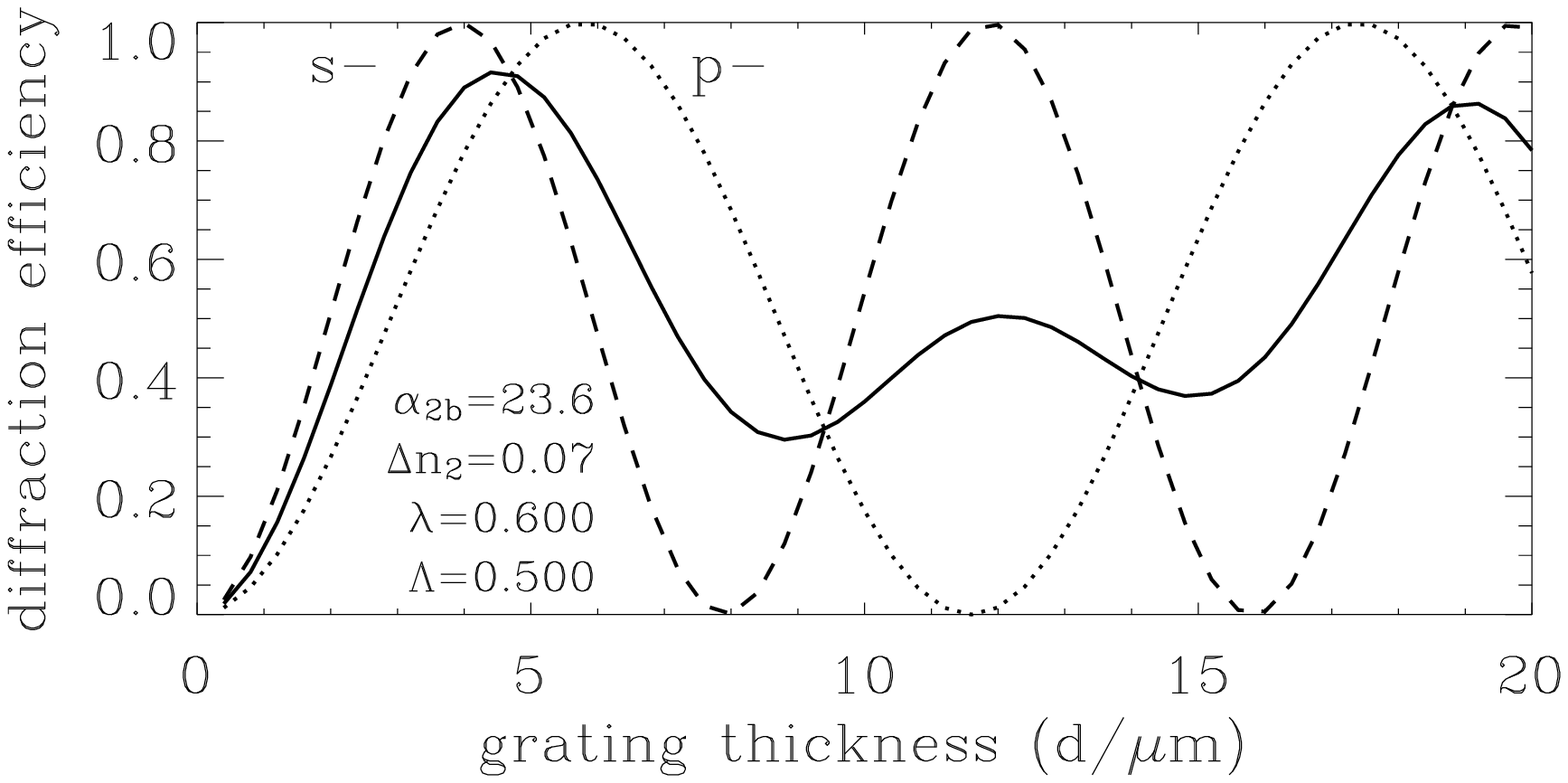}{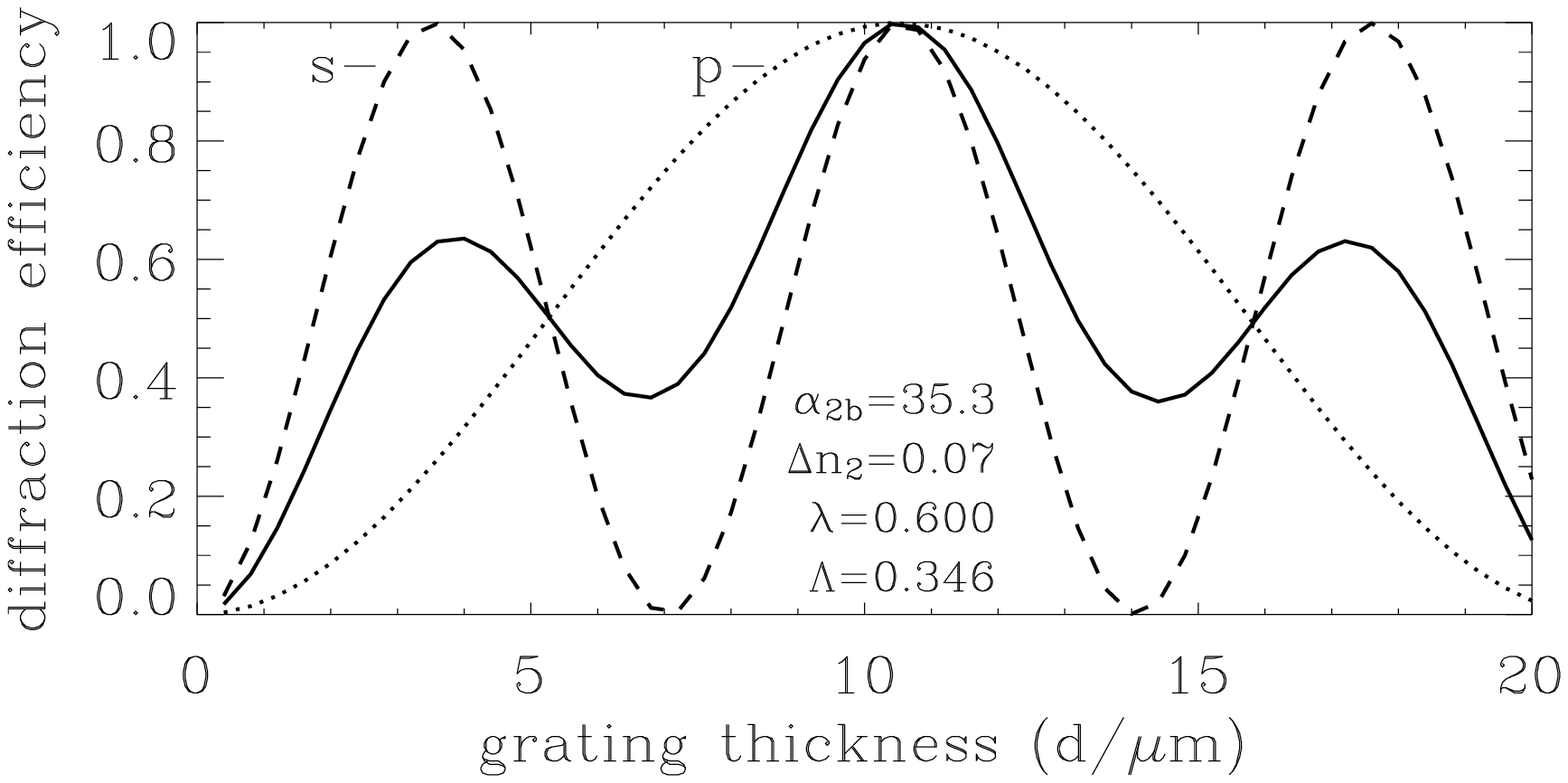}
\caption{Variation of diffraction efficiency versus grating thickness 
  for two different Bragg angles.  The {\em solid lines} represent the
  efficiency of unpolarized light while the {\em dashed} and {\em dotted
    lines} represent the \s- and \p-polarization states, respectively.  Note
  that, in the lower plot, the \s- and \p-states are in phase with a thickness
  of about 10\micr. This special case represents an example of an \spphased\
  grating design.}
\label{fig:eff-thick}
\end{figure}

The Bragg condition is not the only condition for high efficiency.  The
diffraction efficiency depends on the semi-amplitude of the refractive-index
modulation ($\Delta n_2$) and the grating thickness ($d$) in addition to the
incidence and diffracted angles.  \citet{Kogelnik69} determined first-order
diffraction efficiencies at the Bragg condition, using an approximation that
is accurate (to within 1\%) when
\begin{equation}
 \rho = \frac{\lambda^2}{\Lambda^2 \, n_2 \, \Delta n_2} > \rho_{\rm limit}
 \label{eqn:kog-approx}
\end{equation}
where $\rho_{\rm limit}\approx10$.  Substituting $\lambda = 2 \, n_2 \,
\Lambda \, \sin \alpha_{2b}$ (Eqn.~\ref{eqn:bragg} with $m=1$) and rearranging
gives
\begin{equation}
  \sin\alpha_{2b} > \sqrt{\frac{\rho_{\rm limit}}{4}\frac{\Delta n_2}{n_2}} \:.
  \label{eqn:kog-approx2}
\end{equation}
Thus, for a given refractive-index modulation, Kogelnik's theory is accurate
for Bragg angles above a certain value.  For unpolarized light, the Kogelnik
efficiency is given by
\begin{equation}
 \eta = 
 \frac{1}{2} \sin^2 
   \left[ \frac{\pi \, \Delta n_2 \, d}{\lambda \, \cos \alpha_{2b}} \right] 
 + \frac{1}{2} \sin^2 
   \left[ \frac{\pi \, \Delta n_2 \, d}{\lambda \, \cos \alpha_{2b}} 
   \cos(2\alpha_{2b}) \right]
  \label{eqn:kogelnik}
\end{equation}
where the first term is for \s-polarized light (the electric vector is
perpendicular to the fringes) and the second term is for \p-polarized light
(the electric vector is parallel to the fringes).\footnote{Note that in some
  papers the equation for \p-polarization efficiency is incorrectly quoted.
  The additional coupling parameter, the cosine of the sum of the two angles
  [$\cos(2\alpha_{2b})$ in this paper], should be placed {\em within} the
  $\sin^2$ brackets. This can make a significant difference.  If the coupling
  parameter is placed outside the brackets, it implies that the efficiency in
  \p-polarization is always less than in \s-polarization whereas
  Eqn.~\ref{eqn:kogelnik} does not. Instead, the thickness of the grating must
  be larger to produce the same efficiency in \p-polarization. This is
  demonstrated in Fig.~\ref{fig:eff-thick} and is confirmed by coupled-wave
  analysis. Note also that we only consider first-order diffraction in this
  paper because VPH gratings generally have significantly lower efficiencies
  in higher orders.}

Figure~\ref{fig:eff-thick} shows the variation of efficiency versus grating
thickness for two different Bragg angles (with fixed $\Delta n_2=0.07$ and
$\lambda=0.6$\micr).  The efficiencies were determined using {\small
  GSOLVER$^{\rm TM}$},\footnote{GSOLVER Version 4.0, A diffraction grating
  analysis tool (P.O.~Box~353, Allen, TX 75013: Grating Solver Development
  Company), available at {\tt http://www.gsolver.com/} .} which provides a
numerical calculation using rigorous coupled-wave analysis (RCWA)
\citep{MG78,MG81,MG83,GM85}. In these cases, the numerical results are in
excellent agreement with Kogelnik's theory (Eqn.~\ref{eqn:kogelnik}) because
Eqn.~\ref{eqn:kog-approx} is satisfied.  Note that no surface losses were
included. In the upper plot, the first peaks of the \s- and \p-polarizations
are close together and a diffraction efficiency of about 90\% in unpolarized
light can be achieved (with a thickness of 5\micr).  In the lower plot, the
35.3\degr\ Bragg angle is a special case where the second peak of
\s-polarization matches the first peak of the \p-polarization and near 100\%
efficiency can be achieved (with a thickness of about 10\micr).  This special
`\spphased\ grating', also called a `Dickson grating', was noted by
\citet*{DRY94}.

If we consider only the \s-polarization curve in the upper plot of
Fig.~\ref{fig:eff-thick} (dashed line), notice that theoretical 100\%
efficiency occurs with a thickness of 4\micr.  If the thickness is doubled to
8\micr, the efficiency falls to zero.  This is not surprising, since if it
took 4\micr\ of refractive-index-modulated medium to coherently diffract the
light, then twice as much medium causes destructive interference in the
first-order diffraction direction (and instead the light will pass through
without diffraction). At 12\micr, the \s-polarized light is returned to 100\%
diffraction efficiency.  \p-polarized light is characterized by a reduced
coupling with the medium, which means that larger thicknesses are required to
coherently diffract the light depending on the Bragg angle [the
$\cos(2\alpha_{2b})$ factor in Eqn.~\ref{eqn:kogelnik}].

The differences in efficiency between the two polarization states can be used
to produce polarization-selective devices (\citealt*{KKH90}; \citealt{DRY94};
\citealt{Huang94}). In addition, the polarization properties can be used to
determine the average refractive index of the DCG after processing
\citep{RS92,DRY94}. With this technique, measurements of the average index of
highly processed DCG ($\Delta n_2>0.1$) give low values of
$n_2\approx1.25$.\footnote{Low values of the average refractive index, $n_2$,
  imply that voids are formed in the DCG layer during processing
  \citep{CS70,Meyerhofer77}. The exact mechanism is uncertain and the lowest
  achievable index depends on the processing technique.  In addition, the
  index derived using the polarization properties, and assuming Kogelnik's
  theory, may be lower than the `true index' if birefringence is induced in
  the material \citep{Tholl95}.  However, for VPH design purposes, it is in
  any case more appropriate to use the `Kogelnik index', which can be regarded
  as the effective index, determined by Rallison and others, that is needed to
  satisfy the polarization properties of Kogelnik's theory.  Note also that
  $d$ and $\Delta n_2$ in the equations of this paper should be regarded as
  representing the effective thickness and effective index modulation.}
Unprocessed DCG has an index of 1.54. The average index is important since it
determines the air angles ($\alpha_0$) for the special angle gratings, for
example, the 35\degr\ Bragg-angle \spphased\ grating utilizes incidence angles
of 46\degr\ to 48\degr\ in air \citep{RRD03}.

\subsection{Tuning a VPH grating}
\label{sec:tuning}

A VPH grating can be tuned by changing the incidence angle, which changes the
Bragg condition, to optimize the diffraction efficiency for a desired
wavelength.  For example, consider a VPH grating with 1315 \lmm\ 
($\Lambda=0.76$\micr, $n_2=1.5$), at $\alpha_{2b}=16\degr$, the first-order
Bragg condition gives $\lambda=0.63$\micr\ whereas, at $\alpha_{2b}=20\degr$,
the Bragg condition gives $\lambda=0.78$\micr.  The Bragg wavelength is
generally close to but not necessarily the same as the blaze wavelength, which
refers to the wavelength of peak efficiency for a given incidence angle.  Note
that the blaze wavelength does not depend strongly on the shape of the index
modulations, which are presumed to be sinusoidal in the models.  To detect the
blaze wavelength, it may be necessary to change the grating-to-camera angle as
well as the collimator-to-grating angle (see \citealt{bernstein02} for an
alternative approach using a pair of mirrors).  A VPH grating works best at
one incidence angle and one wavelength (the maximal-peak), and even though the
blaze wavelength can be changed by varying the tilt, the peak efficiency is
reduced away from the maximal-peak.  The maximal-peak is determined by the
grating thickness and the refractive-index modulation, as well as the grating
period and spectral order (\S~\ref{sec:eff-1st-ord}).

To illuminate how the Bragg-condition diffraction efficiency varies, we can
rearrange the equation for the Kogelnik efficiency in the following way.
Substituting for $\lambda$ (from Eqn.~\ref{eqn:bragg}) in
Eqn.~\ref{eqn:kogelnik} (and using a trigonometric identity: $2 \sin\alpha
\cos\alpha = \sin2\alpha$), we derive
\begin{equation}
 \eta = 
 \frac{1}{2} \sin^2 \left[ 
   \frac{\pi \, P_{\rm tune}}{\sin(2\alpha_{2b})} 
   \right] 
 + \frac{1}{2} \sin^2 \left[ 
   \frac{\pi \, P_{\rm tune}}{\sin(2 \alpha_{2b})}  \cos(2\alpha_{2b})
   \right]
  \label{eqn:kogelnik2}
\end{equation}
where 
\begin{equation}
 P_{\rm tune} = \frac{\Delta n_2 \, d}{n_2 \, \Lambda} \:. 
 \label{eqn:p-tune}
\end{equation}
Thus, this `tuning parameter', which depends only on the properties of the DCG
layer, determines how the efficiency of a VPH grating varies with Bragg angle.
In other words, all gratings with the same value of \ptune\ have the same
tunability and the same peak efficiency (subject to
Eqn.~\ref{eqn:kog-approx}); while $n_2\,\Lambda$ sets the relationship between
$\lambda$ and $\alpha_{2b}$ (\S~\ref{sec:bragg-angle}); and $\Delta n_2$ and
$d$ can be adjusted, within a certain range, to set the bandwidth
(\S~\ref{sec:bandwidth}).

\begin{figure}
\epsscale{1.15}
\plotone{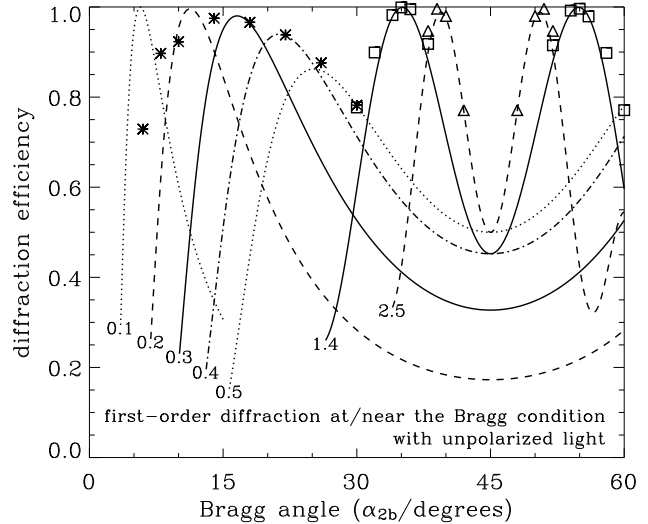}
\caption{First-order diffraction efficiencies versus Bragg angle
  with unpolarized incident light.  The {\em lines} represent efficiencies at
  the Bragg condition determined using Kogelnik's theory for seven gratings
  with \ptune\ values of 0.1--0.5, 1.4 and 2.5 (fixed intrinsic grating
  parameters; see Eqns.~\ref{eqn:kogelnik2} and~\ref{eqn:p-tune}).  The {\em
    symbols} represent maximum designable efficiencies determined using RCWA
  (varying $d$, $\Lambda$; fixed $\lambda=0.6$\micr, $\Delta n_2=0.07$,
  $n_2=1.5$).  The {\em asterisks} represent standard gratings, where the
  optimum thickness with unpolarized light is near the first peak of the
  \s-polarization curve (cf.\ upper plot of Fig.~\ref{fig:eff-thick}, $P_{\rm
    tune} \la 0.5$).  The {\em squares} and {\em triangles} represent
  \spphased\ gratings, where the thicknesses are near the second peak and
  third peak, respectively, of the \s-polarization curve (cf.\ lower plot of
  Fig.~\ref{fig:eff-thick}, $P_{\rm tune} \approx 1.4$; $P_{\rm tune} \approx
  2.45$).  The efficiencies from the RCWA calculations drop at low angles
  because Kogelnik's approximation is no longer accurate below about 20\degr,
  with $\Delta n_2/n_2 \approx 0.05$ (Eqn.~\ref{eqn:kog-approx2}).  Note that
  the grating period, wavelength and Bragg angle are related by
  Eqn.~\ref{eqn:bragg}.}
\label{fig:eff-bragg}
\end{figure}

Figure~\ref{fig:eff-bragg} shows the variation of diffraction efficiency
versus Bragg angle, with unpolarized light, for: (i) fixed gratings, with
various values of \ptune, determined using Kogelnik's theory (represented by
the lines); and (ii) maximum designable efficiencies, with fixed wavelength
and DCG modulation, determined using RCWA (represented by the symbols).  The
asterisks represent standard grating designs while the squares and triangles
represent \spphased\ grating designs. These efficiencies were determined,
using RCWA, by varying $d$ with fixed $\Delta n_2$.\footnote{The optimization
  excluded thicknesses significantly beyond the first peak of the
  \p-polarization curve. With arbitrarily large thicknesses, it is
  theoretically possible to obtain 100\% diffraction efficiency at any angle
  (except 45\degr) with unpolarized light using \spphased\ gratings, e.g.,
  matching the third peak of the \s-curve with the second peak of the
  \p-curve. We do not consider these other \spphased\ gratings because
  increasing the thickness has the disadvantage of reducing the efficiency
  bandwidth (\S~\ref{sec:bandwidth}).}

At angles below about 15\degr, the efficiencies determined using RCWA fall
below that of the maximum efficiencies of curves with \ptune-values\ of
0.1--0.2.  This is because Kogelnik's approximation is no longer accurate
below about 20\degr, with $\Delta n_2/n_2 \approx 0.05$
(Eqn.~\ref{eqn:kog-approx2}). Low-angle efficiency can be increased by
lowering $\Delta n_2$ (and raising $d$).  At angles between 15\degr\ and
30\degr, the maximum efficiencies of curves with \ptune-values\ of 0.3--0.5
closely follow the asterisks showing that Kogelnik's theory agrees with the
RCWA calculations.  This agreement also applies to higher angles.  Around the
angles of 35\degr\ and 55\degr, the designs can make use of special \spphased\ 
gratings that have \ptune-values of 1.3--1.5 (squares).  Around the angles of
39\degr\ and 51\degr, the designs can make use of \spphased\ gratings that
have \ptune-values of 2.4--2.5 (triangles).  Note that at 45\degr, a maximum
efficiency of only 50\% can be achieved because of the loss of \p-polarization
(Eqn.~\ref{eqn:kogelnik}).  To utilize higher Bragg angles, very high
air-to-glass incidence angles are required ($>60\degr$ with $n_2\approx1.25$)
or prisms need to be attached to the grating.

\begin{figure}
\epsscale{1.15}
\plotone{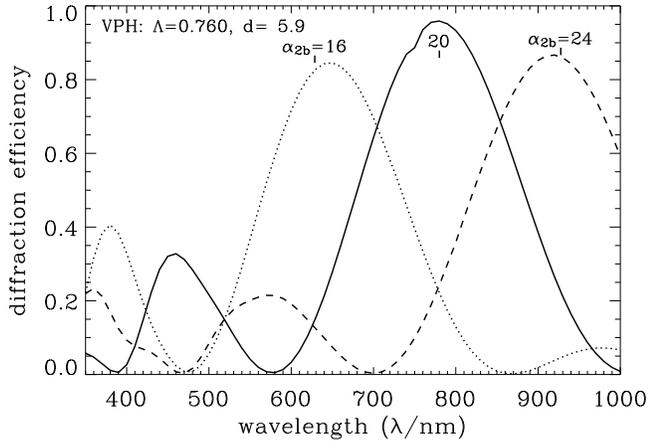}
\caption{Tuning a VPH grating.{} 
  First-order diffraction efficiencies versus wavelength are shown for three
  different incidence angles. The efficiencies were determined using RCWA with
  $\Delta n_2=0.07$, $n_2=1.5$, unslanted fringes and unpolarized light. The
  Bragg angles are shown at the Bragg wavelengths, which are marked by short
  vertical lines. Note that the blaze wavelength (the peak of each efficiency
  curve) is slightly different from the Bragg wavelength for the low and high
  incidence angles.}
\label{fig:tuning}
\end{figure}

The lines in Fig.~\ref{fig:eff-bragg} represent efficiencies derived by
`tuning' various gratings that are optimized for unpolarized light at a
particular angle. At angles between 25\degr\ and 32\degr, it is still possible
to obtain high efficiency ($>90$\%) but with only one linear polarization
state.  Note that the ability of a VPH grating to be tuned in blaze
wavelength, through changing the grating tilt, also results in multi-slit
spectrographs using VPH gratings exhibiting a shift of the blaze wavelength
for objects that are off-axis in the spectral direction \citep{robertson00}.
Figure~\ref{fig:tuning} shows diffraction efficiency versus wavelength for a
VPH grating with 1315 \lmm, tuned to three different blaze wavelengths.  As
the grating tilt is changed, the peak efficiency drops slightly as the blaze
wavelength is moved away from the maximal-peak.

\subsection{Bandwidth of efficiency versus wavelength}
\label{sec:bandwidth}

\begin{figure}
\epsscale{1.15}
\plotone{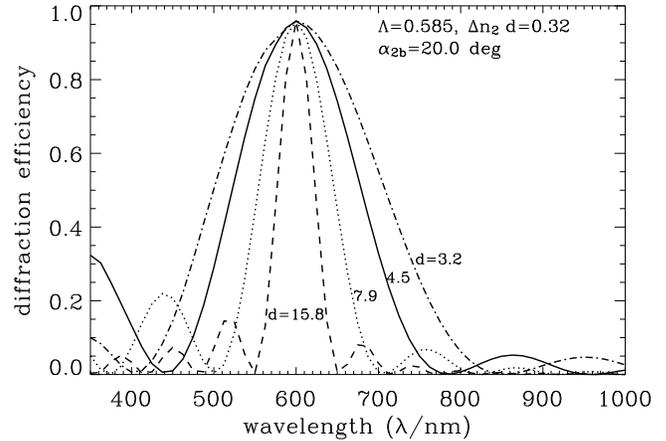}
\caption{Variation of the diffraction efficiency versus wavelength
  for four different grating thicknesses. The efficiencies were determined
  using RCWA.  All the gratings have 1710 \lmm\ unslanted fringes with light
  incident at $\alpha_{2b}=20\degr$ (the first-order Bragg angle for
  $\lambda=0.6$\micr\ and $\Lambda=0.585$\micr).  $\Delta n_2$ is 0.02, 0.04,
  0.07 and 0.10 for each grating, from the thickest to the thinnest,
  respectively.}
\label{fig:bandwidth}
\end{figure}

How does the efficiency decrease away from the Bragg wavelength?  Kogelnik
determined an approximate formula for the full width at half maximum (FWHM) of
the efficiency bandwidth ($\Delta \lambda_{\rm eff}$) in first order, which is
given by
\begin{equation}
\frac{\Delta \lambda_{\rm eff}}{\lambda} \sim 
\frac{\Lambda}{d} \cot \alpha_{2b} \:.
\label{eqn:bandwidth}
\end{equation}
We can see immediately that for a given resolution and wavelength (Bragg angle
$\alpha_{2b}$ and grating period $\Lambda$ fixed), increasing the thickness
decreases the bandwidth.  Figure~\ref{fig:bandwidth} shows efficiency versus
wavelength for four different thicknesses. The efficiencies were determined
using RCWA.

To maximize the bandwidth, $\Delta n_2$ should be as large as possible as long
as the gratings can be manufactured sufficiently thin (remembering that
$\Delta n_2 \, d$ determines the maximal-peak). However, the efficiency
decreases significantly if $\Delta n_2$ is increased such that $\rho$ becomes
much less than 10 (Eqn.~\ref{eqn:kog-approx}). The lost power from first order
diffraction is approximately $1/\rho^2$ \citep{RRD03}.  Therefore, to maximize
the average efficiency across a desired wavelength range, there may be a trade
off between maximizing the bandwidth and maximizing the peak efficiency.  For
Bragg angles greater than 20\degr, this is generally not an issue because the
upper limit on $\Delta n_2$ is set by the manufacturing process.  Values of
$\Delta n_2$ of up to 0.10--0.15 can be achieved using DCG \citep{DRY94}.

\subsection{Other issues}
\label{sec:other-issues}

So far we have dealt mainly with theoretical results either from Kogelnik's
equations or from RCWA numerical calculations.  A number of other points, that
could also affect efficiency, are described briefly below.

\begin{figure}
\epsscale{1.15}
\plotone{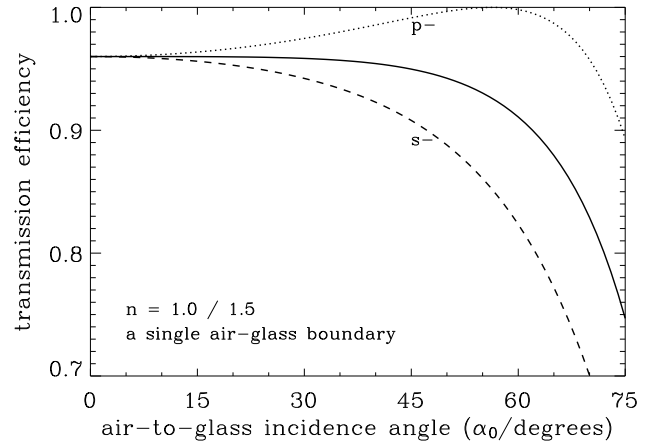}
\caption{Theoretical transmission efficiency at an air-glass boundary.  
  The {\em solid line} represents unpolarized light while the {\em dashed} and
  {\em dotted lines} represent the \s- and \p-polarization states. The
  advantage of \p-polarization for high-resolution spectrographs is apparent.
  The formulae and derivations for these curves can be found in optics
  textbooks, e.g., \citet{Hecht74}.}
\label{fig:air-glass}
\end{figure}

\begin{enumerate}
\item Transmission losses. The main losses will generally be from the
  air-glass boundaries.  Figure~\ref{fig:air-glass} shows transmission
  efficiency versus incidence angle at a boundary.  With $n_0=1$,
  $n_1\simeq1.5$ and no AR coatings, the combined losses will be about 8--10\%
  for incidence angles ($\alpha_0$) from 0--45\degr\ with unpolarized light.
  These can be reduced with AR coatings applied to both surfaces.  For
  wavelengths between 0.4\micr\ and 2\micr, the transmittance of a thin layer
  (5\micr) of DCG is very high and losses are insignificant \citep{BAC98}.
  Between 0.3 and 0.4\micr, losses could be a few percent.
\item The refractive index of the DCG layer. The average index ($n_2$) and the
  semi-amplitude of the modulations ($\Delta n_2$) will not be exactly
  constant.  A small variation with wavelength is not expected to have a
  significant impact on the performance of a VPH grating.  Of possible
  importance is a variation of the index modulation as function of depth or
  position across the surface. In the first case, a reduction with depth means
  that the effective thickness is less than the thickness of the DCG layer,
  and in the second case, the diffraction efficiency will vary with position
  unless there is a high degree of uniformity in the laser beams that produce
  the modulation \citep{RRD03}.  An additional issue concerning the difference
  between non-sinusoidal and sinusoidal refractive-index modulations is
  discussed by \citet{barden00}.  This can improve diffraction efficiency in
  second and higher orders.
\item Defects and errors in manufacturing. Defects could include deviations
  from parallelism between fringes and other non-uniformities across the
  grating.  Errors are deviations of VPH specifications from those requested.
  This may not be important since the grating can be tuned to the required
  wavelength even if the maximal-peak did not meet specifications
  (e.g., \citealt{Glazebrook98nov}).
\end{enumerate}
See \citet{barden00} for the performance evaluation of three VPH gratings for
astronomical spectrographs. 

\section{Example Cases and Discussion}
\label{sec:example-cases}

\subsection{\spphased\ gratings}

In most spectrographs, the polarization states are not separated and therefore
the efficiency with unpolarized light is important. For low resolution
spectrographs ($\alpha_{2b}\la 20\degr$), the theoretical diffraction
efficiency can be above 95\% with standard VPH gratings. At higher resolution,
the efficiency of standard gratings can be significantly lower. Instead,
\spphased\ gratings can be used to obtain higher efficiency at specific
angles (\S\S~\ref{sec:eff-1st-ord}--\ref{sec:tuning}).

\begin{figure}
\epsscale{1.15}
\plotone{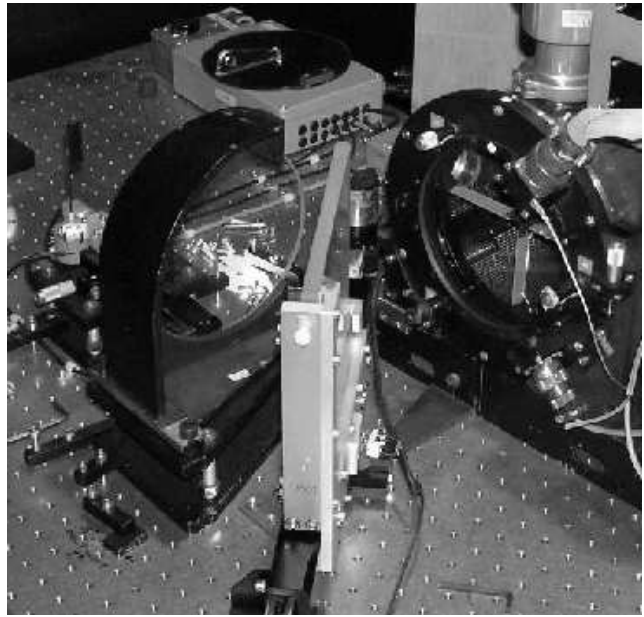}
\caption{The 6dF bench-mounted spectrograph 
  with the 1700 \lmm\ \spphased\ grating in place. There is a 90--100\degr\ 
  beam deviation between the last element of the collimator (left) and the
  camera (right).}
\label{fig:6dF}
\end{figure}

The first \spphased\ (Dickson) grating for astronomy was manufactured by
Ralcon\footnote{Ralcon Development Lab., P.O.~Box~142, Paradise, Utah 84328
  ({\tt http://www.xmission.com/$\sim$ralcon/}), founded by R.~D.\ Rallison.}
for the 6dF multi-object spectrograph at the UK Schmidt Telescope
\citep{saunders01}.  It was specially designed to observe the Calcium triplet
around 0.85\micr\ with a resolving power of about 8000 (first order, 1700
\lmm).  The central wavelength is diffracted with a total beam deviation in
air of 94\degr, which makes use of the 35\degr\ special Bragg angle.  The
theoretical diffraction efficiency is above 95\% in the range
0.835--0.865\micr\ and the performance is near to that.  In addition, the
camera and collimator are close to the grating.  Figure~\ref{fig:6dF} shows
the 6dF spectrograph in this configuration. This has significantly improved
the PSF at the detector in comparison with using reflection gratings that have
similar resolving powers (W.\ Saunders 2003, private communication).

\begin{figure}
\epsscale{1.15}
\plotone{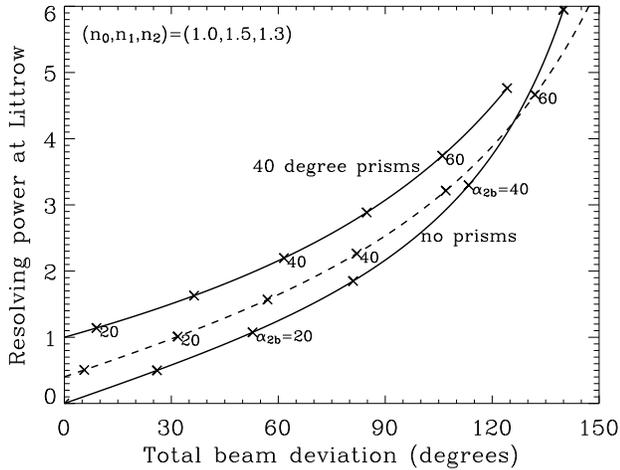}
\caption{Resolving powers of VPH transmission gratings versus total 
  beam deviation. The top line represents a grating immersed between two
  40\degr\ prisms (with $n_1=1.5$), the dashed line between two 20\degr\ 
  prisms and the lower line represents a grating with no prisms attached.  The
  crosses are set at 10\degr\ intervals in Bragg angle (with $n_2=1.3$).  The
  resolving powers are normalized to unity for the zero deviation 40\degr\ 
  prism model. See Fig.~\ref{fig:prism-model} for the prism model and the
  Appendix for the calculation of resolving power. Note that the dispersion
  caused by differential refraction is not included.}
\label{fig:res-beam-dev}
\end{figure}

\begin{figure}
\epsscale{1.15}
\plotone{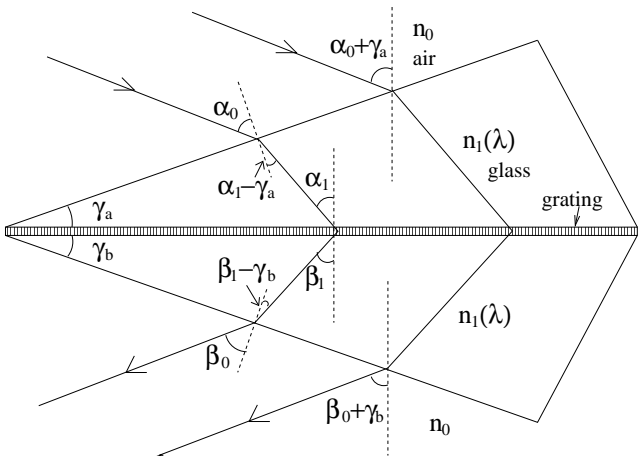}
\caption{Diagram of a prism model for an immersed transmission grating.
  The dependence of the spectral resolving power on the angles and indexes is
  given in the Appendix. In Littrow configuration, with both prism angles
  equal to $\gamma$, the resolving power is approximately proportional to $n_1
  \tan\alpha_1 \cos(\alpha_1-\gamma) / \cos\alpha_0$.}
\label{fig:prism-model}
\end{figure}

If we wish to go to higher spectral resolution but are limited to a certain
maximum deviation between collimator and camera beams, then prisms can be
attached (Figs.~\ref{fig:res-beam-dev}--\ref{fig:prism-model}, Appendix).  For
example, a grating with 2400 \lmm\ and 20\degr\ prisms ($n_2=1.25$, $\Delta
n_2 \, d = 0.73$, $n_1=1.5$) that operates at $\lambda=0.85$\micr\ with a
111\degr\ total beam deviation in air ($\alpha_0=35.5\degr$,
$\alpha_1=43\degr$) can make use of the 55\degr\ special Bragg angle for high
efficiency (\ptune=1.4). This type of grating is challenging to produce
because of the difficulty in testing the efficiency prior to attaching prisms
(consider total internal reflection) but if high resolution and high
efficiency are important over a narrow wavelength range then it could be
useful.  Testing is needed to determine the laser exposure levels for the DCG.
One solution would be to design the grating to work at the 35\degr\ Bragg
angle, which has the same \ptune\ value. This would require testing at a
wavelength related by 0.708 [$=\sin(35.3) / \sin(54.7)$] times the design
wavelength, subject to variations in $n_2$ and $\Delta n_2$ with wavelength
(Eqns.~\ref{eqn:bragg} and~\ref{eqn:p-tune}).  Note that in order for such
gratings to reduce systematic noise in the detected signal, it may be
necessary to use a filter to block scattered light from outside the desired
wavelength range.

\subsection{Separating polarization states in a spectrograph}

If we could envisage an ideal spectrograph, what properties would it have? The
primary problems are scattered light at refractive index boundaries, and the
difficulties of dispersing \s- and \p-polarization states without
compromise of one or the other. We note that both of these problems arise from
the geometry of the wave front with respect to the optical element/grating.
 
A substantial increase in efficiency, perhaps approaching the ideal, could be
achieved by allowing the two polarization states to be handled separately in a
spectrograph. One approach would be to separate the two polarizations at a
polarizing beam splitter \citep{Goodrich91}. They would then propagate along
separate paths where all the optics would be oriented to minimize light loss
for that polarization. In particular, a VPH grating can be optimized for
almost any angle to obtain near 100\% efficiency at blaze wavelength in one
polarization.  An alternative to the use of a beam splitter would be to use a
VPH grating itself.  The collimated beam would encounter a VPH grating in the
normal way, with diffracted light going to a camera. But the grating would be
designed so that first-order diffraction was optimized for a single linear
polarization while zeroth order was optimized for the other
polarization.\footnote{At certain Bragg angles, it is theoretically possible
  to achieve 100\% diffraction efficiency in one linear polarization (\s\ or
  \p), with 0\% diffraction efficiency in the other \citep{DRY94,Huang94}.
  This is analogous to the special angles for \spphased\ gratings.} The
undiffracted transmitted beam could then go on to a second VPH grating
optimized for the other polarization, feeding into a second camera.  Whatever
method is used to separate the polarizations, note that it is not necessary to
achieve complete separation, and hence the requirement for the polarizing
element is less demanding than for a polarimeter.  This is because the only
effect of mixing in a small amount of light with the `wrong' polarization is
that it will be less efficiently processed downstream. But it will still make
some positive contribution to the signal, and the decrease in efficiency will
be a second-order effect.

The cases where separate polarizations would most improve efficiency is with
high spectral resolution. Large beam deviations, and large air-to-glass
incidence angles, are needed to obtain the highest resolutions
(Fig.~\ref{fig:res-beam-dev}). At these large angles, the
boundary-transmission efficiency of \s-polarization is low but the efficiency
in \p-polarization remains high (Fig.~\ref{fig:air-glass}).  For example, one
could design a grating that operates with light incident near Brewster's angle
on the air-glass boundary and with the DCG layer optimized for
$p$-polarization efficiency.  This would provide a high-resolution grating
with near 100\% efficiency in one linear polarization. Naturally, polarimetry
measurements could also take advantage of these one polarization optimized
gratings.  Note also that separating polarization states could be used to
optimize efficiency for reflection grating spectrographs \citep{LA00} and for
background-limited observing during bright Moon phases \citep{BB01}.

We are moving toward precision measurements in many areas of astrophysics. The
major limitation continues to be systematic sources of noise.  In order to
combat this, many experiments are cast as differential measurements, e.g.,
alternating observations of source and background, in order to beat down the
systematic errors. This is a highly effective strategy for dealing with
external noise sources and some internal sources (e.g., apparatus
instability).  However, there are internal sources of noise, which continue to
haunt most spectrographs today, in particular, scattered light. This `ghost
light' is not usually suppressed in differential experiments because it
depends on the distribution of light sources over the field of view. Even with
mitigation strategies based on light baffles and optimized AR coatings, there
is always residual stray light, not least from the optical/IR detector because
of its large refractive index compared to a vacuum.  But this situation is
slowly improving as detectors and matched coatings approach their theoretical
maximum.

The only way to guard against stray light is to consider the role of every
element in the optical train very carefully and to orient the optical elements
accordingly, particularly the choice of AR coating, and orientation of the
interface to the incoming wave front. This is easier to do if the wavefront
has been divided into its \s- and \p-states and each polarization is
considered separately.  One only has to consider the AR coatings in each of
two arms, one for \s- and the other for \p-states, which can be optimized for
throughput.  This is not true for a skew ray in natural light, which would
require a birefringent coating in order to optimize throughput in both
polarization states.

\section{Summary}
\label{sec:summary}

VPH gratings are used in an increasing number of spectrographs because of
their high diffraction efficiency. In this paper, we have outlined the basic
physics necessary to design VPH gratings. In particular: we have defined a
parameter, \ptune, that determines how the efficiency of a grating varies with
Bragg angle; we have described the possibility of creating \spphased\ gratings
that can have 100\% efficiency with unpolarized light at specific angles; and
we have discussed the importance of considering the separate polarization
states. The main points concerning tuning and efficiency are given below.
\begin{enumerate}
\item The grating period ($\Lambda$) and the average refractive index of the
  DCG layer ($n_2$) determine the wavelength as a function of Bragg angle
  (Eqn.~\ref{eqn:bragg} with $m = 1$).
\item The parameter \ptune\ of a grating (Eqn.~\ref{eqn:p-tune}) determines
  how the efficiency varies with Bragg angle (Eqn.~\ref{eqn:kogelnik2}).
  Standard grating designs have $P_{\rm tune} \la 0.5$
  (Fig.~\ref{fig:eff-bragg}).
\item \spphased\ gratings can be created by aligning the peaks of the $s$- and
  $p$-efficiency curves versus DCG thickness at particular Bragg angles
  (Fig.~\ref{fig:eff-thick}).  For example, a grating with $P_{\rm tune}
  \approx 1.4$ has high efficiency with unpolarized light at a Bragg angle of
  35\degr. Here, the second peak of the $s$-curve is aligned with the first
  peak of the $p$-curve.
\item Bragg-condition diffraction efficiencies are lower than predicted by
  Eqn.~\ref{eqn:kogelnik} or~\ref{eqn:kogelnik2} if Kogelnik's condition is
  not satisfied (Eqn.~\ref{eqn:kog-approx} or~\ref{eqn:kog-approx2}).
  However, the efficiency can still be above 90\% as long as $\rho\ga 3$ (the
  lost power is approximately $1/\rho^2$).
\item The FWHM of the efficiency curve is approximately inversely proportional
  to the thickness of the grating (Eqn.~\ref{eqn:bandwidth},
  Fig.~\ref{fig:bandwidth}). Therefore, it is generally optimal to have the
  thinnest possible DCG layer subject to manufacturing limitations and lost
  power from first-order diffraction.
\end{enumerate}

We have shown how VPH gratings can be manufactured and exploited to ensure
higher transmission and better suppression of stray light. This will
necessarily force instrument designers into a smaller parameter space, but we
feel that there is sufficient freedom within that space to account for most
design issues. In any event, we deem these considerations to be paramount if
systematic sources of noise are ever to be effectively removed from the
apparatus.

\acknowledgments{We would like to thank Sam Barden, Chris Clemens, Richard
  Rallison, Will Saunders, and Keith Taylor for information and helpful
  discussions; and we thank the referee for comments, which have improved the
  paper.  Some of this research was funded by a design study for OSIRIS at the
  Gran Telescopio Canarias.}

\appendix

\section{Resolving powers of immersed transmission gratings}

VPH gratings can be sandwiched between glass prisms.  This reduces the total
beam deviation and reduces the air-to-glass incidence angle ($\alpha_0$), for
a given grating and wavelength.  This can be useful because the total beam
deviation is limited by the physical sizes of the camera and collimator and
because higher incidence angles on air-glass boundaries give higher reflection
losses (for unpolarized light). Here, we give the equations for calculating
the resolving power of a transmission grating immersed between two prisms.

Figure~\ref{fig:prism-model} shows the prism model that we are using with the
appropriate angles defined.  Light passing through the prism and the immersed
grating, with a total beam deviation of $\alpha_0 + \beta_0 + \gamma_a +
\gamma_b$, obeys the following equations:
\begin{eqnarray}
n_1 \sin (\alpha_1 - \gamma_a) & \: = \: & n_0 \sin \alpha_0
\label{eqn:imm-grat-entry}
\\
\sin \beta_1 & \: = \: & \frac{m \lambda}{\Lambda_g n_1} - \sin \alpha_1
\\
n_0 \sin \beta_0 & \: = \: & n_1 \sin (\beta_1 - \gamma_b)  \:.
\end{eqnarray}

The resolution can be determined by solving 
\begin{equation}
\beta_0(\alpha_0, \lambda + \Delta \lambda) = 
\beta_0(\alpha_0 - \Delta \alpha, \lambda) \:. 
\label{eqn:res-prism-solv}
\end{equation}
Here, the output angle is regarded as a function of the input angle and the
wavelength.  This expression represents the condition that incrementing the
wavelength by $\Delta \lambda$ shifts the output image by the same amount as
does the change in the incidence angle across the slit width.  $\Delta \alpha$
is the angular size of the slit in the collimated beam and is given by
\begin{equation}
\Delta \alpha = \theta_s \frac{f_{\rm tel}}{f_{\rm coll}}
\label{eqn:sky-angle}
\end{equation}
where $\theta_s$ is the angular size of the slit on the sky, and $f_{\rm tel}$
and $f_{\rm coll}$ are the effective focal lengths of the telescope and
collimator.  If $n_1$ and $n_0$ are independent of wavelength (i.e., ignoring
differential refraction), then Eqn.~\ref{eqn:res-prism-solv} can be solved
analytically to give a resolving power of
\begin{equation}
\frac{\lambda}{\Delta \lambda} = \frac{f_{\rm coll}}{\theta_s f_{\rm tel}} \, 
\frac{n_1}{n_0} \, \frac{\cos (\alpha_1 - \gamma_a)}{\cos (\alpha_0)} 
\left( \tan \alpha_1 + \frac{\sin \beta_1}{\cos \alpha_1} \right) \:.
\end{equation}
Note that, with $\gamma_a = \gamma_b = 0$, this reduces to the well known
equation for the resolution of an unimmersed grating:
\begin{equation}
\frac{\lambda}{\Delta \lambda} = \frac{f_{\rm coll}}{\theta_s f_{\rm tel}} \, 
\left( \tan \alpha_0 + \frac{\sin \beta_0}{\cos \alpha_0} \right) \:.
\end{equation}

To include the dispersive effects of glass ($n_1$ varying with $\lambda$),
Eqn.~\ref{eqn:res-prism-solv} can be solved numerically.  Differential
refraction marginally increases the resolving power for typical VPH grism
designs.

In Littrow configuration, $\gamma_a=\gamma_b\,(=\gamma)$ and
$\alpha_i=\beta_i$, with a total beam deviation of $2\alpha_0 + 2\gamma$, the
resolving power ($n_0$ and $n_1$ constant) is given by
\begin{equation}
\frac{\lambda}{\Delta \lambda} = 
\frac{f_{\rm coll}}{\theta_s f_{\rm tel}} \, \frac{n_1}{n_0} \, 
\frac{\cos (\alpha_1 - \gamma)}{\cos (\alpha_0)} \, 2 \tan \alpha_1  \:.
\label{eqn:res-littrow}
\end{equation}
The usefulness of Littrow configuration is three fold: (i) VPH unslanted
fringes can be used (slanted fringes may curve during DCG processing); (ii)
the beam size remains about the same, which keeps the camera optics smaller
and simpler; and (iii) the angular size of the slit remains nearly the same.
With unslanted fringes, the important Bragg angle is given by
$n_2\sin\alpha_{2b} = n_1\sin\alpha_1$.  Figure~\ref{fig:res-beam-dev} shows
resolving powers at Littrow versus total beam deviation with the Bragg angle
annotated.  Note that for a given grating (fixed diffraction order, wavelength
and \lmm), prisms typically {\em reduce} resolving power, but for a given
total beam deviation, prisms typically {\em increase} resolving power
\citep{Wynne91,LA00}.

\end{document}